\documentclass[conference]{IEEEtran}

\usepackage{cite}

\usepackage[british]{babel}
\usepackage{graphicx}
\usepackage[binary-units]{siunitx}
\usepackage[obeyspaces,hyphens]{url}
\newcommand{\URL}[1]{$\langle$\url{#1}$\rangle$}

\IEEEoverridecommandlockouts
\IEEEpubid{\makebox[\columnwidth]{978-1-4673-9698-1/18/\$31.00~\copyright
2018 IEEE \hfill}
\hspace{\columnsep}\makebox[\columnwidth]{ }}

\begin{document}

\title{Optical TEMPEST}

\author{\IEEEauthorblockN{Joe Loughry}
\IEEEauthorblockA{University of Denver \\
Denver, Colorado 80208 USA \\
Email: joe.loughry@cs.du.edu}}


\maketitle

\begin{abstract}
  Research on optical TEMPEST has moved forward since 2002 when the first
  pair of papers on the subject emerged independently and from widely
  separated locations in the world within a week of each other. Since that
  time, vulnerabilities have evolved along with systems, and several new
  threat vectors have consequently appeared. Although the supply chain
  ecosystem of Ethernet has reduced the vulnerability of billions of devices
  through use of standardised PHY solutions, other recent trends including
  the Internet of Things (IoT) in both industrial settings and the general
  population, High Frequency Trading (HFT) in the financial sector, the
  European General Data Protection Regulation (GDPR), and inexpensive drones
  have made it relevant again for consideration in the design of new products
  for privacy. One of the general principles of security is that
  vulnerabilities, once fixed, sometimes do not stay that way.
\end{abstract}

\section{Introduction}

Since the publication sixteen years ago of the first two papers on optical
TEMPEST, open source information on security of compromising emanations---and
side channels in general---has exploded. Before the late 1990s, only a
handful of papers had been published in the open literature on the subject of
TEMPEST\footnote{TEMPEST, here, is the U.S.\ National Security Agency (NSA)
code name for `the problem of compromising radiation'---radio frequency (RF)
or acoustic radiation, according to the original reference, which was only
declassified in 2007---including both exploitation and control
\cite{NSATempest2007}. Beyond the 1972 definition of TEMPEST, military and
academic researchers have since expanded the spectrum of interest to include
DC, optical, thermal, magnetic, and acceleration side channels.} and the
topic was mostly relegated to the folklore of infosec; the only two
scientific studies of unintentional compromising emanations remained van Eck
\cite{vanEck1985} and Smulders \cite{Smulders1990} although Wright's book
around the same time \cite{Wright1987} described anecdotal reports dating
back to the first world war. But following Kocher's seminal 1996 paper on
side channel attacks \cite{Kocher1996} and Kuhn \& Anderson \cite{Kuhn1998a}
having shown---essentially by adding forward error correction to EMC---that
covert channels were not limited by the system boundary, two papers appeared
within a week of each other on complementary aspects of optical emanations
\cite{Kuhn2002,Loughry2002a}. Since then, these two papers, between them,
have been cited more than 300 times.

\subsection{Organisation}

The first part of this paper is a brief review of the results in the first
paper on optical TEMPEST and a critique of mistakes that were made by one
of the first investigators of the phenomenon, in the methodology, model
building, writing, and follow-up of the earliest research. This is followed
by a survey of ways that later authors have repaired the damage. Finally, a
cautionary
tale of the danger of forgetting lessons learnt in security is rounded out by
a description
of some interesting aspects of the design of a new information security
product with
these principles in mind.

\section{The story I never told you before}

Optical TEMPEST was discovered by one of the junior IT people in a bank. In
the raised-floor
computer room on the sixth floor of a high-rise glass building in downtown
Seattle, surrounded by other glass high-rises---their own computer rooms
visible at night by the reddish glow of Light Emitting Diode (LED)
indicators---I was working very late. Dial-up modems had not yet gone
completely extinct, \SI{10}{\mega\bit\per\second} Ethernet was increasingly
common on PCs, and leased lines ran everywhere from the computer room to
branch offices---thousands of them. This was the environment where optical
TEMPEST was discovered in 1992. After a few nights of data gathering, I told
my postgraduate professor at Seattle University, Dr David Umphress, about it.
Cautious experiments were performed. A literature survey was
quietly done to see if anyone had ever noticed it before, and the National
Computer Security Center (NCSC) was asked if they knew of it. All inquiries
ran into a classified information roadblock; just about the only thing that
was known for sure at the time, in the open literature, was that nearly
everything about TEMPEST was classified (but the name was probably not an
acronym).

\subsection{What we found}

Table \ref{table:optical_tempest_findings} summarises the relevant
experimental findings from our 2002 paper \cite{Loughry2002a}. In brief, we
found that Class III devices leaked all of the information processed by the
device in the form of optical emanations, and more than a third of devices
tested were of this type. LED status indicators are both inexpensive \&
reliable---making them attractive to circuit designers---and fast enough to
respond to high-speed digital signals, which is what makes them hazardous
from an information security perspective. Visual indicators only need to be
fast enough to communicate information to human eyes, but LEDs are much
faster. From an attacker's perspective, the fact that an LED turning on and
off at a rate exceeding about \SI{30}{\hertz} appears to human eyes to be lit
steadily means that information can be transmitted right past observers, the
way dogs can hear ultrasonic sounds inaudible to humans. This has both
accidental and intentional infosec implications.

\begin{table}[!t]
    \renewcommand{\arraystretch}{1.3}
    \caption{Summary of findings from the original optical TEMPEST research
        paper of 2002 (adapted from \cite[Tables I and II]{Loughry2002a}).}
    \label{table:optical_tempest_findings}
    \centering
    \begin{tabular}{|c|c|c|c|}
        \hline
        Class & Correlated with & Risk & How many \\
        \hline
        I & State of the device & Low & 10 percent (4) \\
        II & Activity Level & Medium & 54 percent (21) \\
        III & Data (content) & High & 36 percent (14) \\
        \hline
    \end{tabular}
\end{table}

Class III optical emanations were measured almost entirely coming from data
communication devices, ranging from low cost to very high cost equipment. For
reasons described in the next section, Ethernet devices were mostly
unaffected, though we looked hard to find any that might be. We named the
effect `optical TEMPEST'.

\subsection{Missed opportunities}

Around this time, I made a serious mistake. Unbeknown to me, Markus Kuhn, a
postgraduate student at Cambridge, was working along similar lines
\cite{Kuhn2003}. I had
thought about optical emanations from video display screens but dismissed the
idea as physically impossible, without ever testing it. I was wrong. I
thought the decay time of Cathode Ray Tube (CRT) phosphors was too slow to
carry information about the video signal\footnote{Speaking of video signals,
laser printers intensity-modulate an infrared laser with a video signal, and
some plastics are transparent to infrared light. It might be worthwhile to
measure laser printers for information-bearing optical emanations outside
visible wavelengths, as Kubiak has done for RF
\cite{Kubiak2014a,Kubiak2017b,Kubiak2017c,Kubiak2017d}, Ula\c{s} {\it et al.}\
have done for conducted powerline emanations \cite{Ulas2016}, and Enev {\it
et al.}\ have done for conducted powerline emanations from video displays
\cite{Enev2011}.} in the diffuse light available to non-line-of-sight
interception, and consequently, I never looked for it.

Markus Kuhn found and successfully exploited a tiny ripple near the peak of
the response curve of CRT phosphors. The gross shape of the curve belies the
fact that that tiny ripple is detectable in the time domain optical signal,
if your detector is fast enough.

Kuhn's detector---a photomultiplier tube---was better than mine; the
gain--bandwidth product of a photomultiplier is superior to that of the
large-area
photodiode/transimpedance amplifier combination I used, but we were looking
for different things: Kuhn was looking for diffuse emanations from the entire
screen at once, and
I was looking for for line-of-sight photons from a particular LED on an
isolated piece of
equipment (although we did eventually figure out how to separate multiple
superimposed signals from diffuse optical emanations collected by
non-line-of-sight means). The difference is that Markus Kuhn actually looked,
and he found an effect that I missed.

\subsection{Delay of first publication}

Why didn't we publish in the nineteen-nineties? Part of the reason was
perfectionism; I wanted
to be able to explain the phenomenon and make predictions, not only describe
it. In addition, by that time I had left the bank and was working for a
defence contractor on a classified project. I had a security clearance now,
and a greatly expanded awareness of counterintelligence sources and methods.

And so, following procedures, we submitted the paper to NSA for approval to
publish. It took a year and a half to approve.\footnote{Actually, it didn't
quite happen that way. Our paper was approved for publication very quickly,
in only a few weeks; we submitted it to the 10th USENIX Security Symposium
where it was immediately accepted. A few days later, NSA called us back, and
in a panic, insisted that we withdraw the paper from the conference. I had to
apologise to the programme chair; it was awfully embarrassing, and the delay
in publishing was almost two years.} Eventually, NSA wrote back and said,
`approved for public release'. I wonder what they spent all that time doing.

\subsection{Roads not taken}

Universal Serial Bus (USB) devices hadn't happened yet, but today they're
ubiquitous. Later and informal investigation of some light-up USB cables---a
fad that didn't last---turned up evidence of Class II optical emanations
related to data passing through the USB cable, but evidence for Class III
optical emanations remains inconclusive.

Considered narrowly outside the scope of optical TEMPEST, for the purpose of
this paper, are line-of-sight attacks that essentially reduce to direct or
indirect imaging of the display
\cite{Backes2008,Balzarotti2008,Backes2009a,Raguram2011,Xu2013a,Jenkins2013a}.
This is somewhat unfair, as the original attack by Loughry \& Umphress
required, for the most part, line-of-sight access---except, as previously
mentioned, in \S 8.2 of our original paper---but in the time domain, not
space. Optical TEMPEST is
a time domain effect. Other remote attacks employing optical means, such as
visual or interferometric measurement of keyboard or printer acoustic
emanations \cite{Asonov2004,Zhuang2005,Berger2006,Backes2010}, or a highly
novel reverse covert channel using a document scanner \cite{Nassi2017a}, are
outside the scope of this paper. Screen burn-in, for example, is data
remanence, not optical TEMPEST \cite{MDH1998a}. Finally, induced optical
emanations
\cite{Sepetnitsky2014a,Guri2016b,Guri2017a,Guri2017b,Lopes2017a,Guri2017c,
Zhou2017,Zhou2018a} and \cite[Appendix A]{Loughry2002a} are properly
considered out-of-band covert channels
\cite{Lampson1973,Hanspach2014,Carrara2016}, despite being compromising
optical emanations in the time domain, because they are purposely induced by
a nefarious software or hardware agent, or by activity that is controllable
by a third party, introduced into the target system by the attacker. Control
and remote sensing of magnetic fields, or instantaneous electrical power
demand, similarly are a covert channel, not TEMPEST
\cite{Guri2018b,Guri2018e}.

\section{MAC and PHY}

Of the potential optical signal sources available in a standard office
computer (power light, hard disk activity indicator, keyboard LEDs, network
interface link indicators, charging indicators on laptops, optical disc read
head and activity indicators, scanners, motherboard LEDs, optical mouse, and,
of course, the screen), we
looked hard, at the time, at the other two large populations of blinking
LEDs in the world: disk activity lights\footnote{Guri {\it et al.}\ finally
made it work in 2017 by hovering a drone in the air outside the building
\cite{Guri2017a}; unlike data exfiltration using keyboard LEDs
\cite[Chapter 90]{Stephenson1999}, the hard disk LED channel is covert, not
clandestine.} and Network Interface Cards (NICs).
Neither source proved fruitful; we were unable to find any evidence of Class
III optical emanations from storage devices or link activity indicators on
Ethernet cards. The one exception---and it was a bad one---was WAN interfaces
on the back panel of enterprise routers, devices which live in racks that
sometimes back up to windows; see \cite[\S 4.3.1]{Loughry2002a}. Aside from
those, the complete absence of compromising optical emanations from Ethernet
link activity LEDs is believed to be a consequence of the fact that the
Ethernet protocol is well divided into two layers: MAC and PHY.

In the Ethernet protocol, the Media Access Control layer (MAC) marshals bits
into frames and hands them off to the Physical layer (PHY), which deals
exclusively with voltages and waveforms and wires, or radio, or fibre
optics. The MAC talks to the PHY using a protocol called Media Independent
Interface (MII)---GMII for gigabit Ethernet---over a pair of 4-bit-wide
parallel channels (for sending and receiving) clocked at \SI{25}{\mega\hertz}
\cite{TI2009a}.

In the case of twisted pair wire, only a few suppliers make PHY chips and
generally they do it right, providing dedicated pins on the PHY chip for
connecting LEDs for status indication and internally stretching pulses to the
LEDs to make high-speed activity visible to human eyes. In most PHYs, the
minimum duration of pulse stretching is programmable, and in some PHYs it can
even be turned off \cite[Table 39]{Intel2011a}. The contrast here with the
situation we found regarding relatively low-speed serial interfaces is stark;
there, the temptation was seemingly overwhelming for circuit designers to
drive LEDs directly from generously high voltage and high current serial
communication signals---arguably providing reliable indication of signal
quality and perhaps of marginal signal levels at very low cost. Garden
variety LEDs are plenty fast enough to reproduce amplitude-modulated signals
well into the nanosecond range without any special driver circuits required.

\subsection{High Frequency Trading (HFT)}\label{section:HFT}

Sometimes, security problems that you thought you had fixed already, come
back to bite you.

In the early years of this century, a new style of automatic financial
trading appeared, facilitated by the convergence of gigabit per second
networks, computers with 64-bit address spaces, and deregulation
\cite{Lewis2014a}. Their trading advantage came from the finite speed of
light; by physically locating their trading algorithms as close as possible
to the exchange, they could eke out a response time advantage measured in
milliseconds. With the margin between success and failure so narrow, and
backers willing to spend money on bespoke hardware in return for larger
profits, HFT traders pursued ever-smaller improvements in latency and
responsiveness to changes in market conditions and requirements, culminating
in Field Programmable Gate Array (FPGA) or Application Specific
Integrated Circuit (ASIC) implementation of a minimal gigabit Ethernet MAC,
as fast as physics would allow and additionally capable of three things that
conventional Ethernet hardware could not do:

\begin{enumerate}
    \item ultra-low latency (\si{\micro\second}),
    \item ability to read data transmitted near the beginning of an Ethernet
        packet before the entire packet had been received, and
    \item ability to cancel a speculative trade---if needed---after the
        Ethernet packet had begun transmitting, but before it had
        finished.\footnote{The trick was accomplished by purposely corrupting
        the checksum at the end of a packet, relying upon correct behaviour
        of the exchange's Ethernet interface to discard the packet instead of
        processing it.}
\end{enumerate}

Their systems did not have to be universally interoperable, only compatible
enough to talk to the exchange, and that only for the few months the hardware
was typically used before being replaced by something even faster
\cite{Hurd2018a}.

The risk in this kind of cowboy engineering is that of Chesterton's fence;
non-obvious safeguards may be dropped, leaving the implementation vulnerable
to exploitation. While not described in that reference ({\it ibid.}) it is not
unreasonable to speculate that HFT engineers---there were many HFT groups
besides the one in Korea---may have looked critically at the PHY in their
search for another few microseconds to harvest. And developmental hardware,
especially, sometimes needs monitoring or diagnostic LEDs for debugging. It
is purely speculation, but there might be a `window' of opportunity for rival
HFT firms with a telescope and very high speed photodetector to exploit any
incautiously situated LED indicators connected directly to high-speed
registers.

\section{Design of a new product with optical TEMPEST principles in mind}

Optical TEMPEST began in a bank, took a holiday in the Intelligence Community
(IC), and now has circled back to fintech. The remaining frontier is privacy.

Under the General Data Protection Regulation (GDPR) in Europe, and to a
lesser extent the Health Insurance Portability and Accountability Act of 1996
(HIPAA) in the United States, the privacy of individuals and their personal
information is protected by law. In the U.S.\ at least, this makes health
care providers more risk-sensitive than they are cost-sensitive. One
particular problem---amongst many---that needs to be solved in the U.S.\
arises from a quirk of the U.S.\ Food and Drug Administration (FDA), the main
regulator of medical diagnostic and therapeutic devices. The cost of gaining
FDA approval for use of a medical device is high, necessitating sometimes
years of clinical trials, and extensive design and development
documentation.\footnote{The situation is little different in either
commercial aviation or military and intelligence community systems for
classified information: process maturity, formal or semi-formal design, and
exhaustive testing before certification and approval for use.} As a result,
perhaps millions of vulnerable medical devices---only a few years old---exist
that never have got the required security patches for their embedded
computers' operating system (OS) as recommended by the OS manufacturer. The
reason for the shortfall in software maintenance is the excessive cost of FDA
recertification in the event any diagnostic- or therapeutic-relevant changes
are made to the configuration of a medical device \cite{Talbot2012}. In fact,
the analogous situation happens in classified military and intelligence
community networks as well. Commercial aviation experiences the problem less
than either classified networks or healthcare for two reasons: firstly, being
mobile, aviation control systems tend to be more isolated and special-purpose
embedded computers than the Commercial Off-the-Shelf (COTS) hardware favoured
by medical device and intelligence community developers; and secondly, DO-178
\cite{DO-178C}.

In this section I describe some of the design considerations for new hardware
development informed by experience with optical TEMPEST vulnerabilities and
countermeasures. The notional infosec product described is intended to
isolate vulnerable medical devices with the aim of protecting individuals'
privacy by eliminating a potential mode of entry of hackers to the hospital's
internal computer networks.

In a parallel universe to the HFT hardware designers in the previous section,
we use similar techniques to different ends; the MAC here is a state machine
implemented in hardware, not for low-latency but for high-security; the PHY
returns, for security reasons---in the interest of complete transparency of
implementation---to its roots in the magnetics of IEEE 802.3i, where the
number of turns in a toroidal transformer can be counted by hand. The design
and development methodology is that of the intelligence community, but the
anticipated buyer does not reside in the U.S., and is not expected
particularly to trust the U.S.\ government. The only reasonable defence
against this level of mistrust is believed to be complete openness and
transparency of design, development, and implementation.

\subsection{Simplicity and transparency of implementation}

Part of the design is essentially an optoisolator for the purpose of domain
separation between the `private' and `public' sides of the privacy problem.
Integrated circuit optoisolators can be bought but they are designed for
galvanic isolation, not infosec.

\begin{figure}[!t]
    \centering
	\includegraphics[height=2in]{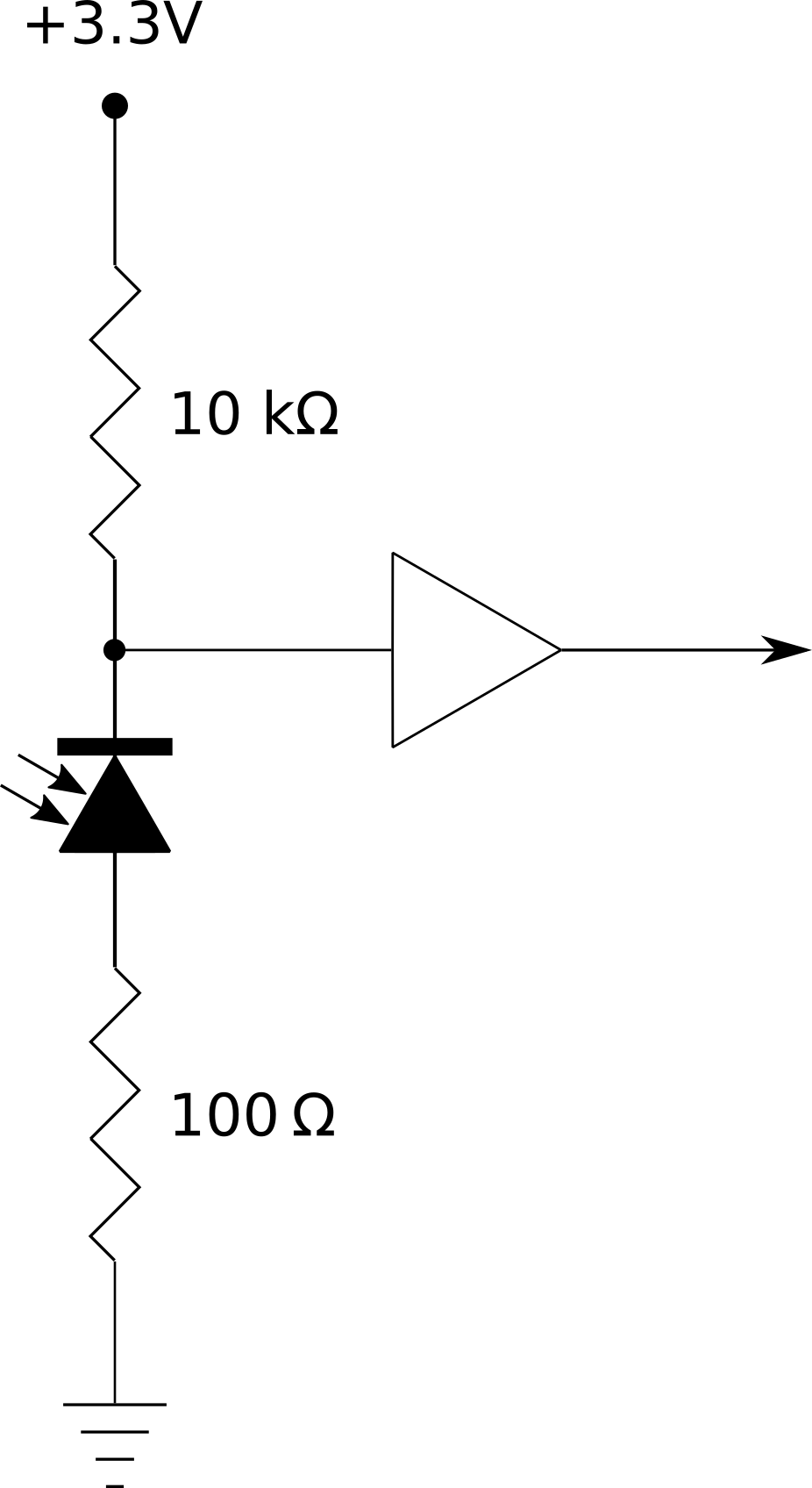}
	\caption{The photodiode circuit is purposely made as simple as possible
        for complete transparency---the goal was not simplicity or parts
        count--of implementation; the \SI{10}{\kilo\ohm} pull-up resistor is
        for reliability together with a \SI{100}{\ohm} series resistor to
        protect the bidirectional driver (here represented by a generic
        buffer) from being shorted to ground in case it were accidentally set
        to output a HIGH logic level at the same time the photodiode were
        illuminated.}
	\label{figure:photodiode_pullup}
\end{figure}

The receiver side of the optoisolator circuit is shown schematically in
Figure \ref{figure:photodiode_pullup}. The photodiode operates in reverse
bias ({\it i.e.}, photoconductive) mode for two reasons: firstly, speed, and
secondly, simplicity of implementation. The same photodiode operated in
photovoltaic mode would be more sensitive to very low level signals and have
a lower dark current, but would require a transimpedance amplifier for
current-to-voltage conversion, which would make the design more complicated
and thereby more difficult to evaluate for security. There is no amplifier
stage because electronics are relatively more difficult to verify than
optics. Not shown are three additional layers of protection, implemented by
passive and active optical, mechanical, and physical means between emitter
and detector to ensure that information flows only in one direction.

\begin{figure*}[!t]
    \centering
	\includegraphics[width=\textwidth]{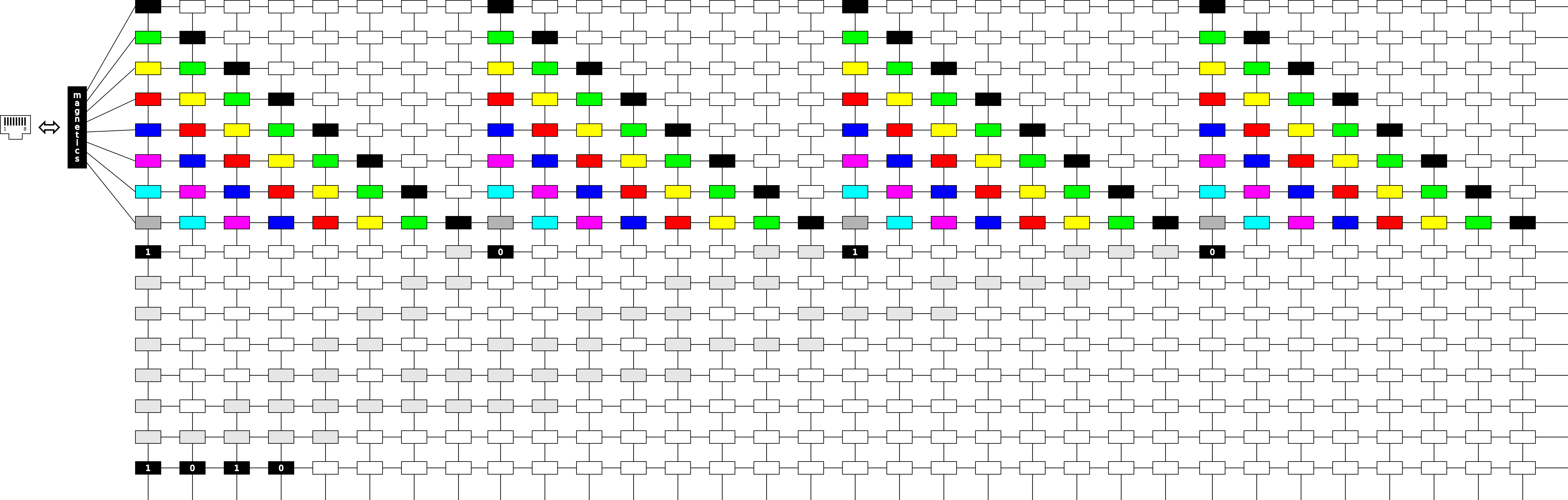}
	\caption{Deep pipeline implementation of an Ethernet MAC; only a small
        portion of the pipeline is shown. The MII is clocked asynchronously
        by the state machine as bits are de-marshalled efficiently into one
        slot each. Profligate expenditure of resources trades off for very
        favourable parallelism and equally transparent computation of sizes,
        offsets, padding, checksum, and digital signature application and
        validation.}
	\label{figure:deep_pipeline}
\end{figure*}

The rest of the design is equally unconventional (Figure
\ref{figure:deep_pipeline}). Rather than using available `IP' blocks
(Intellectual Property---pre-designed software modules delivering standard
functionality such as ARM cores or Ethernet MAC in a chip), we prefer a
clean-room
design approach built up from IEEE 802.3 standards using semi-formal methods,
avoiding closed-source IP. The result, we believe, in combination with open
tests, will be considered trustworthy by everyone in the world.

\subsection{Energy Gap}

The gold standard of information isolation is the `air gap'---physically
separating a system containing sensitive information from attackers. But as
Clive Robinson and others have pointed out, in the era of wireless
networking, air gapping is insufficient.\footnote{Of course, Stuxnet proved
that air gaps can fail in other ways, because they must necessarily be
bridged by sneakernet \cite{Kushner2013a,Langner2013a}.} He coined the term
`energy gap' to include both physical communication channels---RF, acoustic,
optical, thermal, magnetic, or acceleration---in addition to more analogical
interpretations of `energy' such as the work factor required to break
encryption. The design of our notional product puts principles of physics,
not algorithms or mathematics, between the attacker and his target, in an
attempt to widen the energy gap. Optical TEMPEST risks are sealed up in
machined metal cavities; mechanical interlocks supplement logical ones.

\subsection{Design discussion}

Simplicity of implementation is paramount. There are no amplifiers, no MAC
or PHY chips, no process nodes that cannot be de-capped and have a
representative sample to be examined under a microscope. The development tool
chain must be open source and international.

It must be acknowledged that optical TEMPEST, ironically, is a vulnerability
that can be exacerbated by extreme reliance on simplicity of implementation,
as happened with LED indicators on data communication equipment from about
1968--1998.
But experience with optical TEMPEST and side channels has inspired so many
other researchers to find and exploit diverse vulnerabilities, that the only
remaining avenue of approach is to strike as many components as possible from
the system, in the belief that a component that is not there cannot fail, has
no vulnerabilities, and lasts forever.

\section{Conclusion}

Since the publication in 2002 of the first peer-reviewed research on
compromising optical emanations, other researchers have carried it further.
But technological progress has shifted the boundaries of what was possible,
necessitating re-visit of the same vulnerabilities from time to time. This is
a general principle of security; vulnerabilities, once fixed, sometimes do
not stay fixed.

\section*{Acknowledgements}

Thanks to the anonymous reviewers of EMC Europe 2018 and to the many
anonymous and pseudonymous members of the communities of Hacker News
\URL{https://news.ycombinator.com} and Bruce Schneier's blog
\URL{https://www.schneier.com/} for helping develop and critically examine
many of the ideas in this paper.


\IEEEtriggeratref{35}


\bibliographystyle{IEEEtran}
\bibliography{IEEEabrv,consolidated_bibtex_file}

\end{document}